\documentclass[%
 reprint,
 amsmath,amssymb,
 aps,
twocolumn,prl]{revtex4}
\usepackage{amssymb}
\usepackage{amsmath}
\usepackage{graphicx}					
\usepackage{epstopdf}					
\usepackage[latin1]{inputenc}				
\usepackage[T1]{fontenc}					

\usepackage{version}				
\usepackage{cases}						

\usepackage{pstricks}					

\usepackage{mathrsfs}
\usepackage[squaren, Gray]{SIunits}			

\usepackage{color}
\usepackage{psfrag}
\usepackage{enumerate}
\usepackage{amssymb}
\usepackage{amsmath}
\usepackage{wasysym}
\usepackage{hyperref}
\usepackage{color}

\begin{document}

\title{Stochastic chaos in a turbulent swirling flow}

\author{D. Faranda}
\affiliation{%
LSCE-IPSL, CEA Saclay l'Orme des Merisiers, CNRS UMR 8212 CEA-CNRS-UVSQ, Universit\'e Paris-Saclay, 91191 Gif-sur-Yvette, France
}

 \author{Y. Sato}
\affiliation{ RIES/Department of Mathematics, Hokkaido University, N20 W10, Kita-ku, Sapporo, Hokkaido 001-0020, Japan }
\altaffiliation{London Mathematical Laboratory, 14 Buckingham Street, London, WC2N 6DF, UK}

\author{B. Saint-Michel}
\affiliation{Univ Lyon, Ens de Lyon, Univ Claude Bernard, CNRS, Laboratoire de Physique, F-69342 Lyon, France}

\author{ C. Wiertel, V. Padilla, B. Dubrulle,  F. Daviaud,}
 \email{francois.daviaud@cea.fr}
\affiliation{%
SPEC, CEA, CNRS, Universit\'e Paris-Saclay, CEA Saclay
91191 Gif sur Yvette cedex, France
}

\pacs{Valid PACS appear here}

\begin{abstract}
We report  the  experimental evidence of   the existence of a random attractor  in a fully developed turbulent swirling flow. By defining a global observable which tracks the asymmetry in the flux of angular momentum imparted to the flow, we can first reconstruct the associated turbulent attractor and then follow its route towards chaos. We further show that the experimental attractor can be modeled by stochastic Duffing equations, that  match the quantitative properties of the experimental flow, namely the number of quasi-stationary states and transition rates among them, the effective dimensions, and the continuity of the first Lyapunov exponents. Such properties can neither be recovered  using deterministic models nor  using stochastic differential equations based on effective potentials obtained by inverting the probability distributions of the experimental global observables.  Our findings open the way to low dimensional modeling of systems featuring  a large number of degrees of freedom and  multiple quasi-stationary states.  
\end{abstract}

\maketitle

In the absence of external forcing, the only stationary state of a viscous flow is the trivial zero-velocity state. This state obviously respects all the symmetries of the system. Subject to a given forcing of intensity $\mu$, the trivial state can reach a non-trivial steady state (SS), the characteristics of which depend on  $\mu$ and symmetry properties of the forcing. For  $\mu$ below a critical value $\mu_c$, the SS is time independent, and respects all the symmetries of the forcing compatible with the boundary conditions. As $\mu$ is increased past $\mu_c$, the SS gradually breaks all the forcing symmetries, resulting in fluid motion switching from time-independent to periodic, chaotic, and ultimately reach -- at $\mu=\mu_T\gg\mu_c$  -- a turbulent state in which fluid motion is extremely irregular. This state however recovers all the symmetries of the forcing and the system in a statistical sense~\cite{frisch1995turbulence}. 
The turbulent flow is characterized by a dynamics with a large number of degrees of freedom, resulting from the wide range between the length scale at which energy is injected  and the scale at which it is dissipated. This motivated  Landau~\cite{landau1944problem}  to describe it as a quasi-periodic state,-i.e. the superposition of a growing number of modes with incommensurate oscillation frequencies, resulting from an infinite number of bifurcations with increasing $\mu$. This picture was challenged by Ruelle and Takens~~\cite{ruelle1971nature} who proved that  turbulent states are in general not quasi-periodic and conjectured that they could be described by a small number of degrees of freedom, i.e. by a low dimensional "strange attractor"~\cite{ruelle1971nature}  on which all turbulent motions concentrates in a suitable phase space. This conjecture was fueled  by seminal studies on prototype flows such a Taylor-Couette~\cite{brandstater1987strange} or Rayleigh-B\'enard convection~\cite{berge1984order,ahlers2006experiments}, where it was shown that the transition to turbulence actually follows the traditional roads to \textit{deterministic chaos} via the appearance of two or three characteristic frequencies and either quasi periodicity with frequency locking, period doubling or intermittency~\cite{manneville1995dissipative}. However, it was soon realized that this paradigm only survives during the transition to turbulence.  For $\mu_c <\mu \ll \mu_T$, all tentatives~\cite{grassberger1986do,nicolis1984there,lorenz1991dimension} to find the strange attractor of a turbulence state failed. Does this mean that we must abandon all hope to apply tools from dynamical systems theory to turbulent flow?

We provide  experimental evidence that the answer is negative  using  a laboratory model experiment in highly turbulent conditions. The key idea is that even if a turbulent flow is characterized by a large number of degrees of freedom, some of them are less important than others, and can be lumped into a \textit{noise} term with a few relevant parameters. This motivates the shift towards the notion of \textit{random dynamical systems}~\cite{lin2008shear,chekroun2011stochastic} and \textit{stochastic chaos}~\cite{lovejoy1998stochastic}.  We are left with the problem of identification of the relevant variables  which represent the main properties of the steady state, in a statistical sense. Since the bifurcation  in this system is connected with symmetry breaking, it is natural to choose this parameter so that it gives  information about the symmetries of the turbulent steady state, in analogy with usual order parameters in statistical physics. In our experiment, the order parameter is a global quantity measuring the response of the flow to an asymmetry of the flux of angular momentum. As the asymmetry is varied, the turbulent state becomes unsteady,and the formerly stable random attractor becomes unstable, in a sequence reminiscent of the topology changes of the  Duffing attractor~\cite{kovacic2011duffing} with varying forcing amplitude.

We use the experimental set-up as described in~\cite{saint2013evidence}. Turbulence is generated in a vertical cylinder of length $H=180$ mm and radius $R=100$ mm filled with water, and stirred by two coaxial, counter-rotating impellers providing energy and momentum flux at the upper and lower end of the cylinder. The impellers are made of disks of radius $0.925 R$, fitted with 8 curved blades.  They are driven in the scooping direction by two independent motors, operating in conditions such that the torques $C_1$ and $C_2$ applied by the flow onto the top and bottom impellers are constant. This procedure guarantees a stationary flux of angular momentum at each impeller. To quantify the global response of the flow to this forcing, we measure independently the rotating frequency $f_1$ and $f_2$ of the two impellers. With a typical mean applied torque $C=(C_1+C_2)/2=1.40$ Nm, we measure  typical mean frequencies $f$ of the order of  $(f_1+f_2)/2=10$ Hz. Our experiment being thermalized at a temperature $T\approx 20$ $^\circ$C, this corresponds  to a typical Reynolds number $Re=2\pi f R^2/\nu=3\times 10^5$, far from the estimated critical Reynolds number for turbulence onset~\cite{ravelet2008supercritical}: $Re_T=3500$.


In the sequel, we consider only data obtained for {\sl fixed } values of the torques $C_i$. This means that there is only one free parameter characterizing the forcing. Due to the symmetry of our experimental set-up,  statistical-mechanical arguments~\cite{thalabard2015statistical} suggest  the choice $\gamma(t)=(C_1(t)-C_2(t))/(C_1(t)+C_2(t))$ as the control parameter. From now on, the parameter $\gamma$, corresponding to the $\mu$ parameter in our model, is understood as a time averaged value of $\gamma(t)$  as $\gamma(t)$ experiences fluctuations in response to the turbulent flow. The amplitude of the fluctuations -- measured as the standard deviation of $\gamma(t)$ -- is substantially independent of $\gamma$~\footnote{See Supplemental Material at [URL will be inserted by publisher] for additional figures.}. We will see that the stochastic behavior of $\gamma(t)$ is the key to the concept of   random attractors. When $\gamma=0$ the top and the bottom impeller are exchangeable, and we have checked that the turbulent state statistically follows this symmetry. As a result, the top and bottoms rotating frequencies are statistically equal: the variable $\theta(t)=(f_1(t)-f_2(t))/(f_1(t)+f_2(t))$ fluctuates around zero and characterizes the symmetries of the turbulent flow~\cite{saint2013evidence,saint2014zero}. 

\begin{figure}
\includegraphics[width=0.50\textwidth]{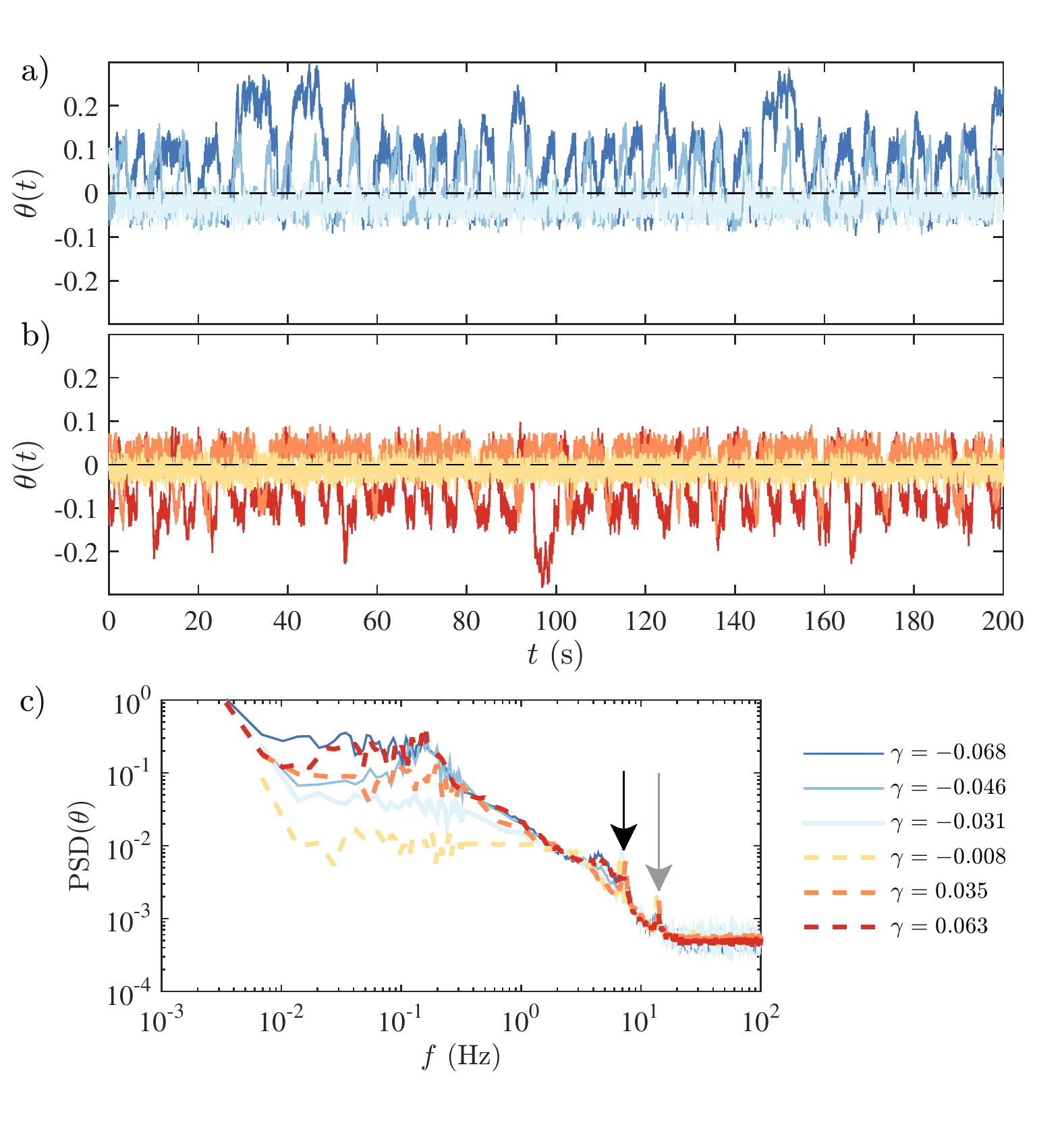}
\caption{ (a-b) Time series  of the global observable $\theta$ reflecting the symmetries of the von K\'arm\'an turbulent flow experiment with torque forcing at $Re=3\cdot10^5$. The dynamics of $\theta(t)$ for $\gamma\simeq 0$ consist of fluctuations around an average value. Instead, for higher values of $|\gamma|$, an irregular switching between different quasi-stationary states appear. (c) Power spectral density (PSD) of $\theta(t)$ for different values of $\gamma$. The black and grey arrows highlight the two spectral peaks $f_0/2$ and $3f_0/2$, where $f_0=14$ Hz  is twice the average rotation frequency of the impellers.}
\label{timeseries}
\end{figure}

The time series and the power spectral density (PSD) of the variable $\theta(t)$ for 6 different values of the parameter $\gamma$ are plotted in Figure \ref{timeseries}. At $\gamma\approx 0$, the time series has the signature of decorrelated white noise, as evidenced by the flat spectrum until $f_G\approx 3$ Hz. At larger values of $f$ the spectrum steepens into an approximate $f^{-1}$ spectrum, superposed with two characteristic frequencies 7 Hz and 10.5 Hz, corresponding to $f_0/2$ and $3f_0/2$,  $f_0= (\bar{f_1}+\bar{f_2})=14$ Hz being twice the average impeller rotation frequency. The spectrum then saturates to white noise  for frequencies larger than $f_N=11$ Hz. 
For other values of $\gamma$, the behavior is identical, with a shift of $f_G$ towards smaller and smaller values. We first eliminate the irrelevant small scale degrees of freedom at $f\geq f_N$ by performing a moving average of the time series over a time window of 12 Hz. The corresponding time series is then analyzed using the standard embedding procedure~\cite{packard1980geometry}, by extraction of the maxima $\theta_m$ (or minima since the results do not change significantly) under the condition that subsequent maxima  cannot fall within 10 Hz.
Once the series of partial maxima is obtained, the attractor is visualized by plotting in a $n$-dimensional phase space, $\theta_m$,  $\theta_{m+1}$, ..., $\theta_{m+n}$. The value of $n$, known as the embedding dimension, plays a crucial role  in the  applications of dynamical systems theory to real data~\cite{kantz2004nonlinear}.

\begin{figure}
\includegraphics[width=.5\textwidth]{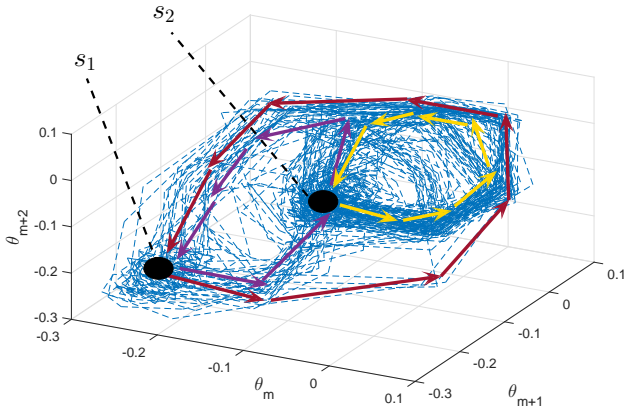}
\caption{von K\'arm\'an turbulent flow attractor  reconstructed for the experiment performed at $\gamma=0.067$. The attractor is obtained by embedding the peaks $\theta_m$ extracted from the time series  $\theta(t)$ in a three dimensional space.  One can identify two quasi-stationary states (labeled as $s_1$ and $s_2$) and three cycles (highlighted by colored arrows). The direction of the arrows indicates the only possible paths to switch from one state to the other. See also supplementary video.}
\label{states}
\end{figure}

\begin{figure*}
\includegraphics[width=1\textwidth]{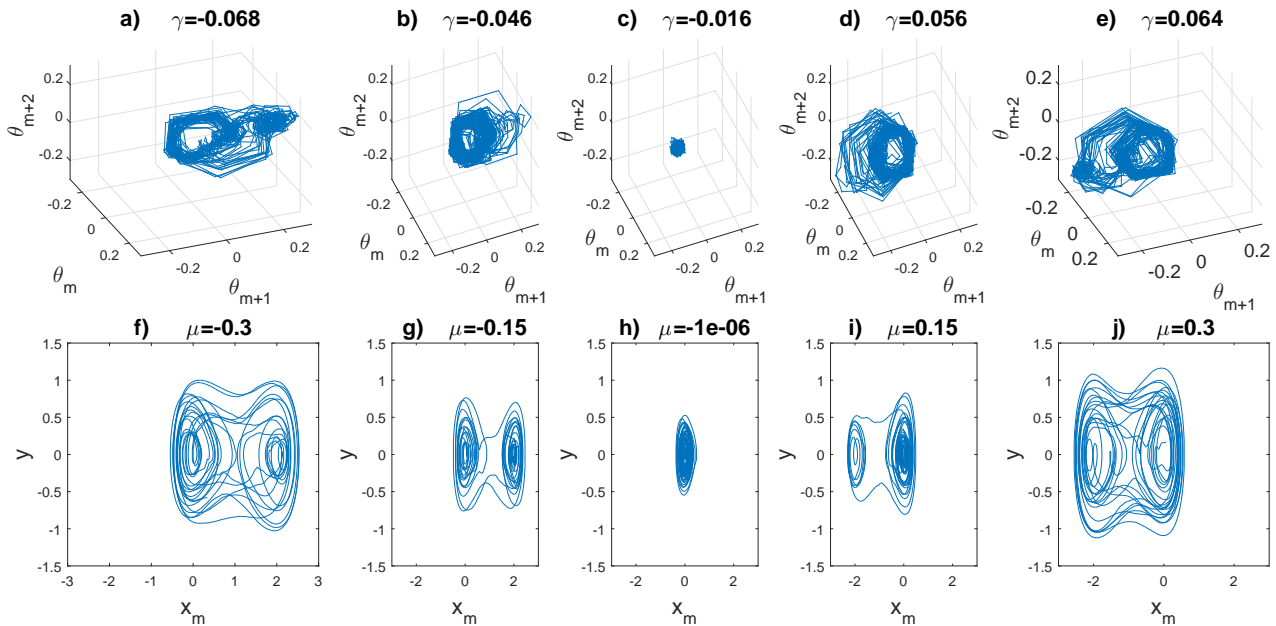}
\caption{(a-e) von K\'arm\'an turbulent flow attractors for 5 different $\gamma$ values. From (c), the sequence shows how the attractor bifurcates symmetrically  into a noisy periodic motion (b,d) and into a noisy chaotic attractor (a,e). (f-j) the same bifurcation sequence reproduced in the stochastic Duffing attractors for 5 different $\mu$ values.}
\label{attractors}
\end{figure*}

\begin{figure}
\includegraphics[width=0.5\textwidth]{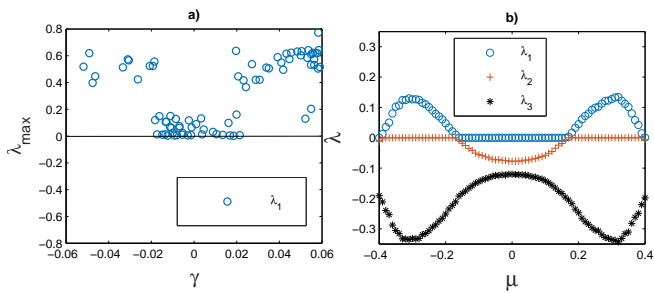}
\caption{(a) First Lyapunov Exponents $\lambda_{1}$  measured for  different $\gamma$ in the von K\'arm\'an turbulent flow (blue circles). For $\gamma\leq 0.02$, $\lambda_{1}\simeq 0$ corresponds to the experiments whose attractor is a  random point attractor. For larger $|\gamma|$, $\lambda_{1}> 0$ and noisy attractors are found. (b) First,  second, and third Lyapunov exponents for different $\mu$ measured in the stochastic Duffing equation with $\sigma=0.2$. The behavior of $\lambda_{1}$ is qualitatively similar to that of the experiments. See  Supplemental Material for the first, the second, and the third Lyapunov exponents at different $\sigma$.}
\label{Lyap}
\end{figure}

The   embedding with $n=3$ for the experimental data with  $\gamma=0.067$ is represented in Figure \ref{states}  (see supplementary video for an animation of the dynamics). This attractor features two quasi-stationary states, $s_1$ and $s_2$. The transitions from one state to another  always follow one of the three cycles highlighted by the arrows, which indicate the only possible ways the system can switch.\\
The dynamics of $\theta$ show a rich bifurcation diagram when $\gamma$ is varied. Some examples are shown in Figure \ref{attractors} (a-e). For $|\gamma|\sim 0$ the attractor is a random point attractor (Figure \ref{attractors}-c). This is the analog of the trivial zero-velocity state of unforced viscous flow. For $  0.02<|\gamma|< 0.04 $ a noisy periodic motion first appears (Figure \ref{attractors}-b,d). For $|\gamma|>0.06$, this  attractor bifurcates into a noisy chaotic attractor (Figure \ref{attractors}-a,e).

We now look now for the minimal dynamical system model capable of representing the dynamics of the experiment. The dynamics of $\theta(t)$ can be mimicked by an autonomous oscillator at frequency $f_0$, while the dynamics of $\gamma(t)$ induced by the turbulent fluctuations, is represented by a stochastic force.  Moreover, the $\theta \to -\theta$ symmetry excludes the presence of a quadratic non-linearity. The simplest model having such characteristics and capable of describing the sequence of bifurcations of the order parameter of  the von K\'arm\'an flow $\theta(t)$ is
the stochastic Duffing equations, a non-autonomous dynamical system with two variables $x$ and $y=\dot{x}$ with random forcing $z$ obeying:
\begin{eqnarray}
dx&=&ydt\nonumber\\
dy&=&(-ay+x-x^3+z\sin\omega t)dt\nonumber\\
dz&=&-\phi(z-\mu)dt +\sigma dW_t,
\label{eq:model}
\end{eqnarray}
where $a=0.2$, $\phi=0.9$, $\sigma=0.2$, $\omega=1$ and $W_t$ is a Wiener process~\cite{ito1974diffusion}. In our model $x$ corresponds to the experimental $\theta$ and $\omega$ corresponds to $f_0$, while $z$ corresponds to the stochastic dynamics of $\gamma(t)$ modeled by the Ornstein-Uhlenbeck process, the simplest stochastic model representing fluctuating dynamics of the control parameter: $\mu$ corresponds to $\gamma$ i.e. the time average of $\gamma(t)$,  $\sigma$ to the amplitude of the fluctuation of $\gamma(t)$, and $\phi^{-1}$  to the characteristic time needed by the system to restore the average $\gamma$~[30]. Using this model, we can generate artificial time series for $x,y$  and  reconstruct the attractor of the stochastic Duffing equations for different values of  the control parameter $\mu$.  Due to the symmetry $\theta \to -\theta$, we have two distinct Duffing attractors for the positive and negative values of $\gamma$. By observing that the quasi-stationary states of Eq. \ref{eq:model} are obtained for $x_s=\pm 1$, the two branches are recovered by shifting $x$ to $x_m=\mbox{sign}(\mu)(x-1)$.   In Figure \ref{attractors} (f-j) we show the stochastic Duffing  attractors in terms of $(x_m,y)$ for different values of $\mu$. As in the turbulent experimental system, there is a bifurcation from a random point attractor to  random periodic attractors, and to  random strange attractors. \\

 To check more quantitatively the analogy between the experimental system and the model, we have computed in both cases the effective dimension with the method proposed by Cao et al.~\cite{cao1997practical}, comparing the experimental data $\theta(t)$ with time series of $x(t)$ of  the same length, for which we have repeated exactly the embedding procedure used for the von K\'arm\'an experiments. We  obtain the effective dimension $n_{\mbox{eff}}\simeq 10$ from the experimental data and $n_{\mbox{eff}}\simeq 9$ from the model. Moreover, the first Lyapunov exponent $\lambda_{1}$ from the data  (Figure \ref{Lyap}-a) and the first Lyapunov exponents computed from the model (Figure \ref{Lyap}-b) show a qualitatively similar behavior  as a function of the control parameters $\gamma$ and $\mu$. The stochastic behavior,  induced by the fluctuation of the control parameter, is essential to get the full bifurcation diagram: by changing $\mu$ in the stochastic Duffing attractor, we observe smooth changes of the Lyapunov exponents, compatible with those observed in the experiments.  In the deterministic case [30] the Lyapunov exponents exhibit discontinuous jumps for increasing $|\mu|$, so that the bifurcation diagram is incompatible with the one observed experimentally. 
 


An essential feature of our reconstruction is the combination of tools from classical dynamical system with ideas borrowed from stochastic modeling, where the influence of neglected degrees of freedom (here the small scales) are described through a noise. The resulting model is a dynamical system with fluctuating control parameter. Such fluctuations  strongly modify the bifurcation diagram of the original system, smoothing the variation of the Lyapunov exponents, in agreement with experimental findings~[30]. As a result, the fluctuations of the order parameter, the transition rates and the bifurcation structure respect the features experimentally observed. Therefore, the random dynamical systems framework is more suitable than the classical dynamical systems to describe our turbulent data. The noise by itself is however not a sufficient ingredient to reconstruct the full system dynamics. Inspection of the turbulent attractor in Fig. \ref{states} may naively suggest that our
 system follows nothing else that a generalized Langevin model, described by the stochastic differential equations (SDEs), like in other turbulent systems ~\cite{benzi1981mechanism,brown2007large}. In such approach, the effective potential may be found by inverting the probability distribution of the global observable (as measured e.g. in~\cite{saint2013evidence} ) to obtain the effective potential describing the fixed points. The transition between quasi-stationary states is then captured by the addition of a noise term, representing the interactions with smaller scales. This approach gives the stationary states with fluctuations but hardly return the correct transition rates as shown by~\cite{lucarini2012bistable}. In fact, the implicit assumption that the potential is one-dimensional (e.g. by taking an overdamped limit) leaves only one possible transition path between  quasi-stationary states. The turbulent attractor provided in Fig. \ref{states} immediately suggests that this description is false: there is more than one path for the  switch between $s_1$ and $s_2$, and the system dynamics cannot be reduced to a single SDE. So, while a SDE is superior to a classical deterministic model~\cite{brown2007large,da1997qualitatively} it fails to reproduce the exact dynamics in the phase space, as  described by Lyapunov exponents and the transition rates. The random attractor model thus appears as the only candidate able to describe both statistical and dynamical features of our data.

We have provided experimental evidence that it is possible to describe 
the large scale motion of a fully-developed turbulent flow with a \textit{random dynamical systems} model with few degrees of freedom, if an appropriate observable reflecting  the flow symmetry is selected. We claim that the large embedding dimensions which prevented the applications of dynamical systems theory to turbulence arise from small scale disturbances which can be modeled in terms of stochastic perturbations. This general picture reconciles the Landau~\cite{landau1944problem} and Ruelle-Takens~\cite{ruelle1971nature} descriptions of turbulence, the former being valid at small scales, and the latter describing the large scale motions. Our findings may be extended to other systems where chaos with large degrees of freedom plays a role, thereby defining the procedure to find attractors in geophysical fluid dynamics~\cite{nicolis1984there,gober1992dimension,lorenz1991dimension,grassberger1983characterization,grassberger1986do}. Like in our turbulent experiments, general oceanic or atmospheric circulations are characterized by general symmetry properties and small scale dynamics that are possibly decorrelated. The main challenge is then to identify the relevant global observable that reflects the system symmetry and that can be used as an order parameter. Indeed, for  some well chosen atmospheric circulation index  (see e.g.~\cite{ambaum2014nonlinear}), the Duffing equation emerges as the minimal model for the description of mid-latitude circulation dynamics.

\bibliography{stochatt}%


\section{Acknowledgments}
 The research leading to these results has  been partially funded by the ERC grant No 338965-A2C2 and the Grant-in-Aid for Scientific Research (C) No. 24540390, JSPS, Japan, and London Mathematical Laboratory External Fellowship, UK. We acknowledge N. Moloney for useful discussion and comments.\\
  
Correspondence and requests for materials
should be addressed to F. Daviaud.~(email: francois.daviaud@cea.fr).


\end{document}